\documentclass[12pt]{iopart}

\usepackage{iopams}
  \expandafter\let\csname equation*\endcsname\relax
  \expandafter\let\csname endequation*\endcsname\relax

	\usepackage{amsmath}
	\usepackage{mathtools}
	\usepackage{graphicx, xcolor}  

\begin{document}

\title{The Gardner correlation length scale in glasses}

\author{M J Godfrey$^1$ and M A Moore$^1$}

\address{$^1$ Department of Physics and Astronomy, University of Manchester,
Manchester M13 9PL, United Kingdom}

\date{\today}

\begin{abstract}
  The Gardner length scale $\xi$ is the correlation length in the
  vicinity of the Gardner transition, which is a transition
  in glasses where the phase space of the glassy phase fractures into
  smaller sub-basins on experimental time scales.  We argue that $\xi$
  grows like $\sqrt{B_{\infty}/G_{\infty}}$, where $B_{\infty}$ is the
  bulk modulus and $G_{\infty}$ is the shear modulus, both measured in
  the high-frequency limit of the glassy state.  We suggest that $\xi$
  might be inferred from stress-stress correlation functions, which is
  more practical for experimental investigation than studying two
  copies of the system, which can only be done in numerical
  simulations.  Our arguments are illustrated by explicit calculations
  for a system of disks moving in a narrow channel, which is solved
  exactly by transfer matrix techniques.
\end{abstract}

\maketitle

\section{Introduction}
\label{intro}
The Gardner transition \cite{gardner:85} is a transition to a state of
full replica symmetry breaking (FRSB) from a state with one-step
replica symmetry breaking: it exists in exotic spin glass models such
as the Potts spin glass \cite{gross:85} or the $p$-spin glass model
\cite{gardner:85}, at least when studied in the mean-field
approximation.  The recent revival of interest in it is because
structural glasses seem to have within their glassy state features
similar to the loss of ergodicity expected at the Gardner transition;
for a review, see \cite{berthier:19}.  To date, most of the numerical
studies of it have focused on finding similarities between the
features observed in dimensions $d =2$ or $d =3$ and those predicted
by the mean-field theory of replica symmetry breaking, which is
certain to be valid only in infinite dimensions
\cite{scalliet:19a,artiaco:19,seoane:18b,liao:19}.  This has involved
looking at the overlaps of the states in copies $A$ and $B$ of the
glassy system, which at some initial time have the particles in the
same positions, but with different initial velocities.  While this
device is useful in numerical simulations for detecting the onset of
non-ergodicity, it is not easy to mimic in a real experiment. 

The Gardner transition in finite dimensions is in the universality class of the Almeida-Thouless transition of spin glassses \cite{Urbani:15}.  According to some \cite{Moore:11, wang:18, Mattsson:95}, the Almeida-Thouless transition is at best an ``avoided" transition for all dimensions $d \le 6$. If so, the Gardner
transition will also be an ``avoided'' transition for dimensions $d \le 6$, in that
the Gardner correlation length can only grow large for hard or soft spheres or
disks \cite{scalliet:19a,scalliet:19b,scalliet:19c,liao:19} and there will be no actual divergence of
this length scale as at a real transition or as found in the
mean-field limit, which is exact only as $ d \to \infty$.

In this paper, we shall obtain a simple expression for the Gardner
length scale which explains in a quantitative way the circumstances in
which it can become large.  Furthermore, we shall show that the length scale can be
obtained by \textit{experiments} on shear-shear or stress-stress
correlations, and does not require the study of two copies of the
system, $A$ and $B$.  Finally we show that at densities above that of
the Gardner transition there are \textit{structural} changes in the
system that are not described by the mean-field calculations.

In the ``state-following'' Gardner transition in structural glasses
one studies the transition within a given glass state, that is, on
time scales short compared to the alpha relaxation time so that the
atoms will not wander far from their initial positions.  This
restriction on the motion of the atoms allows us to derive an
effective elastic model for the glass.  We find that the Gardner
length scale $\xi$ varies approximately as
$\sqrt{B_{\infty}/G_{\infty}}$, where the bulk compressibility
$B_{\infty}$ and the shear modulus $G_{\infty}$ are the values these
elastic moduli take at high frequencies or on time scales less than
the alpha relaxation times.  If the ratio of the moduli becomes large,
the length scale $\xi$ becomes large.  This ratio does indeed become
large at the $J$-point of the glass \cite{epitome,ikeda:15} and so
$\xi$ is large if the state is ``close'' to the $J$-point
\cite{LiuNagel,Ozawa:17c,scalliet:19b,scalliet:19c}; otherwise it will be small.  We shall
illustrate our results by explicit calculations for a system of disks
moving in a narrow channel, whose thermodynamic properties can be
obtained via the transfer matrix technique.  We believe
that our results are generic and  give a simple argument that
suggests that they should extend to both two and three dimensional
glasses.

In Sec.~\ref{model} we describe the simple model system of disks in a narrow channel which we  use to motivate our arguments. In Sec.~\ref{epsilondep} we briefly discuss how our results vary with the width of the channel (but only over a restricted range). The determination of the Gardner length scale from the strain-strain correlation length is given in Sec.~\ref{strains} and we derive a simple effective Hamiltonian which describes its behavior. Finally in Sec.~\ref{discussion} we discuss the implication of our results for hard and soft disks in higher dimensions.

\section{The model: disks in a narrow channel}
\label{model}

\begin{figure}
  \includegraphics[width=\columnwidth]{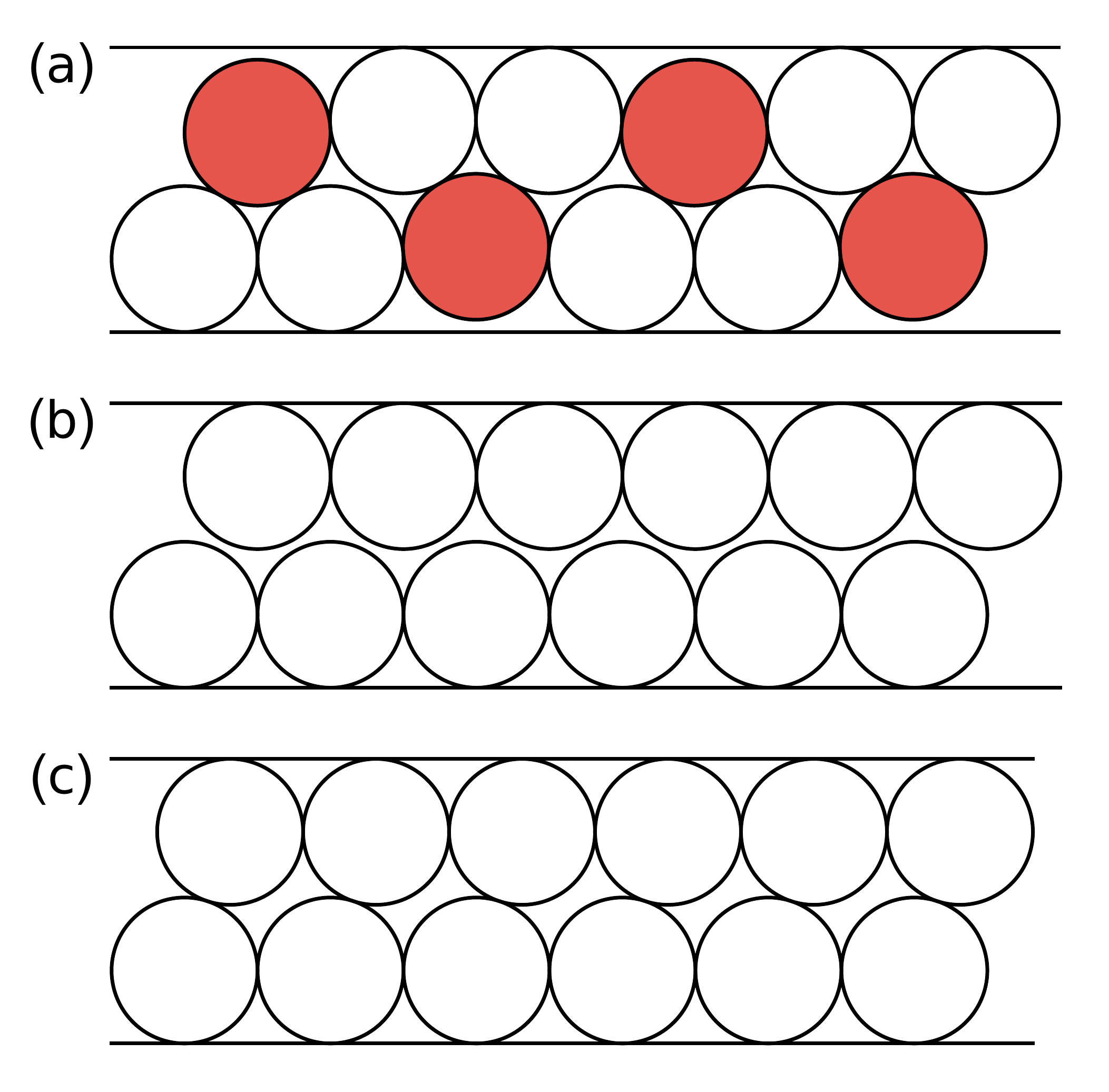}
  \caption{(a) Configuration of disks at the maximum possible packing
    fraction $\phi_{\rm{max}} \simeq 0.8074$ when $h=0.95 \sigma$.
    The red disks do not touch the sides of the channel.  (b)
    Configuration of the disks at a density $\phi_K \simeq 0.8055$.
    This is the highest density state which can be reached when all
    the disks touch a channel wall.  Note that the disks in the upper
    row can be translated with respect to those in the lower row, as
    shown in~(c).}
  \label{fig:configs}
\end{figure}

\begin{figure}
  \includegraphics[width=\columnwidth]{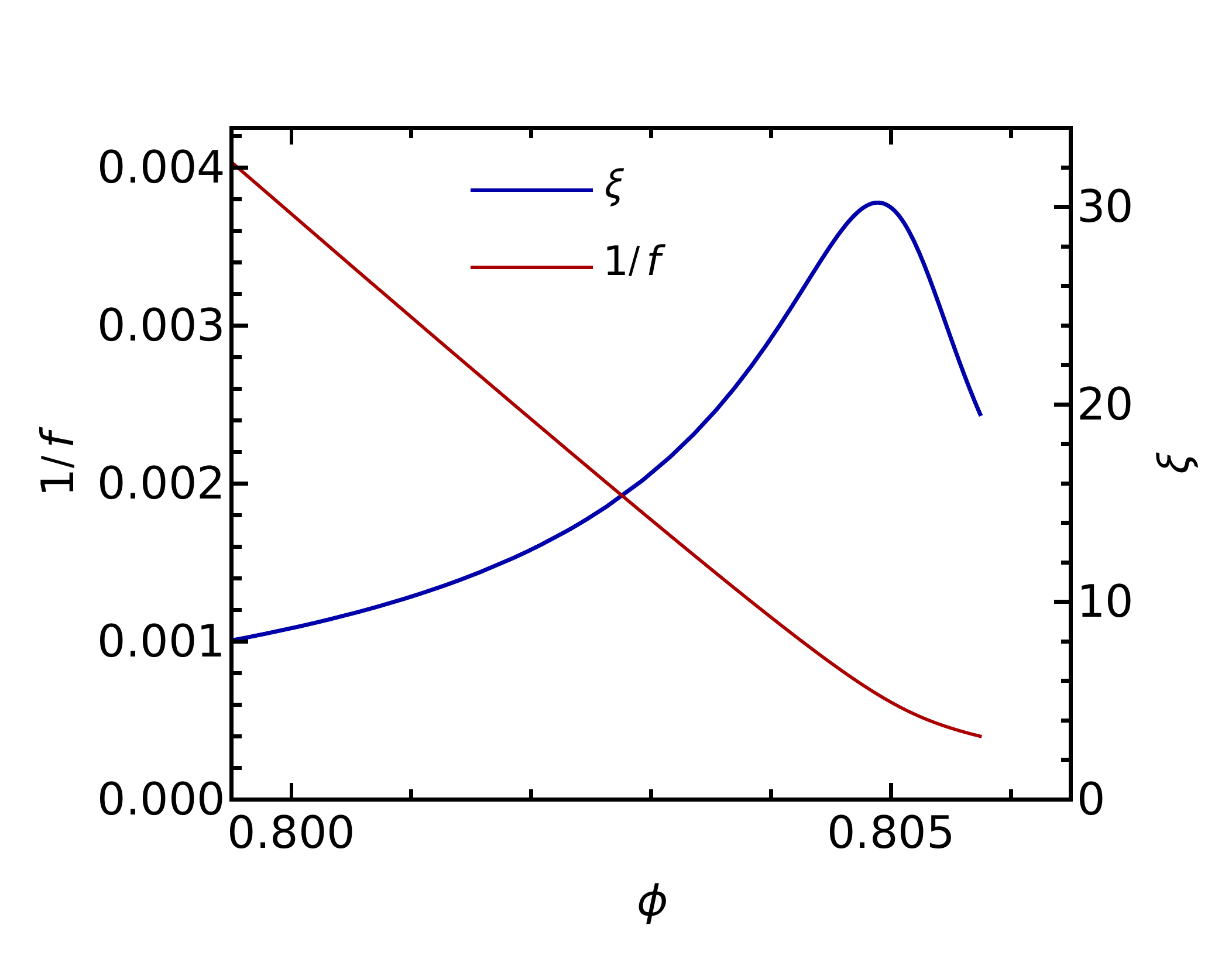}
  \caption{Red: Plot of $f^{-1}$ where $f \equiv \beta F \sigma$
    against packing fraction $\phi$ for disks in a channel of width $h
    =0.95 \sigma$.
    Blue: On the right axis we have plotted $\xi$.  It peaks at the
    packing fraction $\phi_G (\approx 0.8049)$, which is where the
    equation of state curve starts to deviate significantly from a
    straight line.}
  \label{fig:eqnstateh95}
\end{figure}

The system of disks in a narrow channel has previously been studied by
ourselves and others in some detail
\cite{Robinson:16,Godfrey:15,Godfrey:14,bowles:06,Yamchi:12,Ashwin:13,
  Yamchi:15,Kofke:93,
  Varga:11,Gurin:13,charbonneau:18d,hicks:18,Huerta:19}.  Being
effectively one dimensional, it does not have any true phase
transitions.  It has many typical glass features, such as relaxation
times that grow rapidly with increasing packing fraction,
\cite{Robinson:16,Godfrey:14}, the remnants of Kauzmann behavior
\cite{Godfrey:15}, and an avoided Gardner transition \cite{hicks:18}.
Examples of some configurations of the disks are given in
Fig.~\ref{fig:configs}. The channel width available to the centers of
the disks is defined as $h=H_d-\sigma$, where $\sigma$ is the diameter
of a disk and $H_d$ is the width of the channel; we parametrize $h$ by
$h=\sqrt3\sigma/2+\epsilon\sigma$, where $0 < \epsilon <
(1-\sqrt{3}/2)$, so that the disks cannot pass each other.  The
packing fraction $\phi$ is defined as $\phi=N\pi\sigma^2/(4 H_d L)$,
where $N$ is the number of disks in a channel of length $L$. The
center of the $n$th disk has co-ordinates $(x_n,y_n)$, where $y_n$ is
measured from a line down the center of the channel. Thus a disk which
touches a channel wall will have $y_n = \pm h/2$.  The transfer matrix
has been used to obtain numerically exact values for the thermodynamic
properties and correlation functions of the model \cite{Godfrey:15}.
The Gardner correlation length $\xi$ has already been determined from
the eigenvalues of the transfer matrix \cite{Godfrey:15,hicks:18} for
one value of the channel width, $h=0.95 \sigma$.  It peaks at a value
of $\xi \approx 30$ at a packing fraction $\phi_G \approx 0.8049$ for
$h=0.95 \sigma$ (see Fig.~\ref{fig:eqnstateh95}).  However, by
decreasing the channel width ($\epsilon \to 0$) we can make the
Gardner length scale $\xi$ as large as we wish at the avoided
transition, so that we can adjust the extent to which the transition
is avoided, as demonstrated in Sec. \ref{epsilondep}.  In the simulations in two and three
dimensions it has proved to be difficult to determine the Gardner length scale for
hard disks or spheres \cite{liao:19}. 

The dynamics in this system start to slow as ``zigzag'' order sets in
above a packing fraction $\phi =\phi_d\approx 0.48$
\cite{Godfrey:14,Godfrey:15,Robinson:16} for $h =0.95 \sigma$, which is the width mostly used in this paper except in Sec.~\ref{epsilondep}.  Zigzag
order is characterized by successive values of $y_i$ taking opposite
signs and is a form of bond orientational order.  The zigzag order can
be interrupted by defects where successive $y_i$ are of the same sign;
the correlation length $\xi_2$ for zigzag order is approximately half
the average distance between these defects~\cite{Godfrey:14}.  These
defects play an important role in the dynamics of the system.  The
spacing between these defects increases rapidly with increasing
packing fraction $\phi$, such that for $h=0.95 \sigma$, $\xi_2$ passes
$2000$ at $\phi=0.7206$ and reaches $\xi_2=2.3\times 10^6$ at
$\phi=0.76$.  In this study we focus on even higher packing fractions
so that our systems have essentially perfect zigzag order.

\begin{figure}
  \includegraphics[width=\columnwidth]{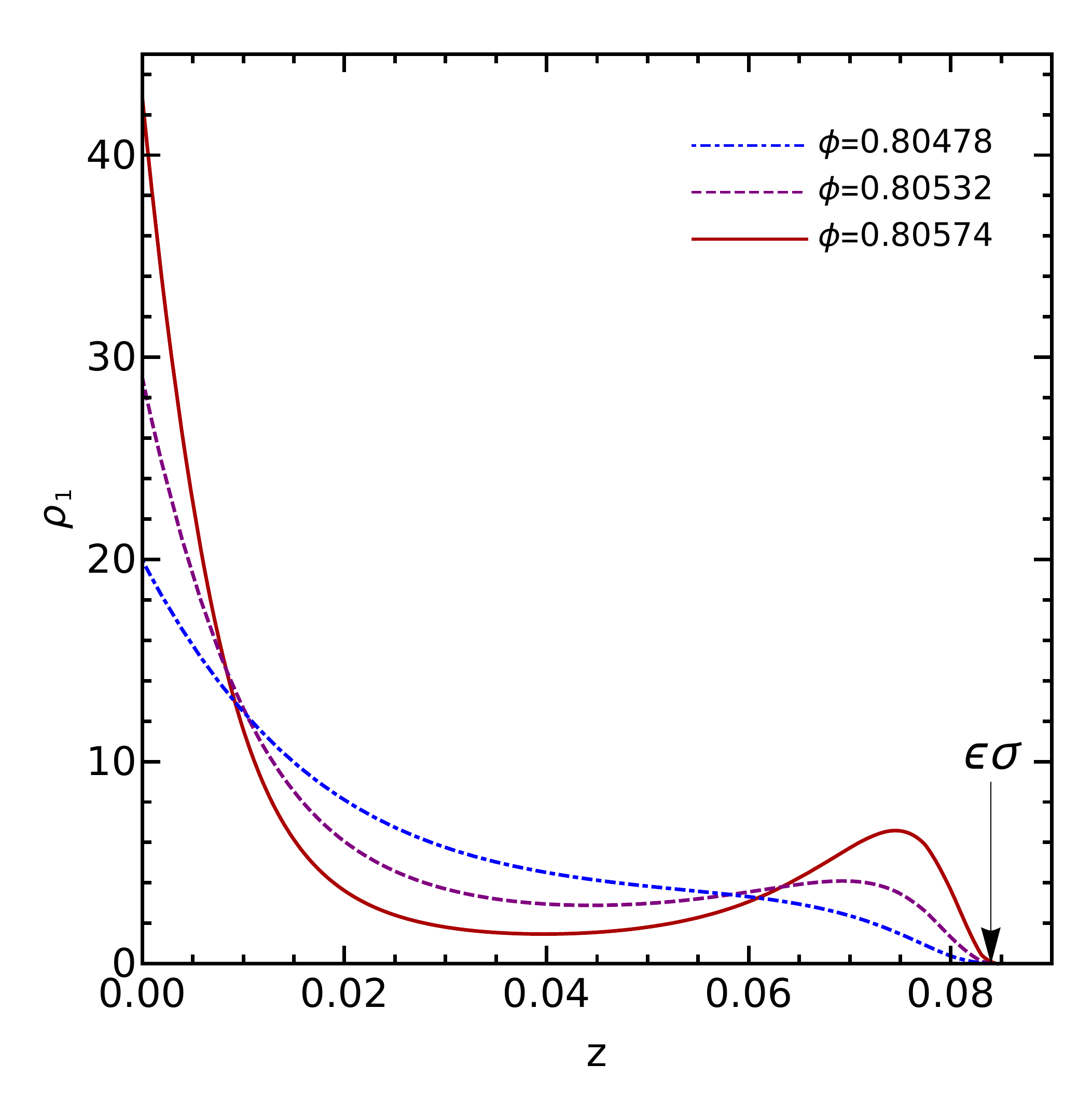}
  \caption{Plot of the density $\rho_1(z)$, where $z=h/2-y$ is the
    distance of a disk from the wall  for three different values of the packing fraction $\phi$ when the channel width $h =0.95 \sigma$. As the packing fraction increases past $\phi_G
    \approx 0.8049$, there is an increasing probability to find disks
    away from the wall.  An arrow marks $z=\epsilon\sigma$, which
    corresponds to the positions of the shaded disks in
    Fig.~\ref{fig:configs}(a).}
  \label{fig:density}
\end{figure}

We used the transfer matrix to calculate the equation of state of the
system \cite{Godfrey:15}.  This is the relation between the force $F$
which has to be applied to pistons at both ends of the system to
confine the $N$ particles within it so that the system's length $L$
corresponds to a packing fraction $\phi$.  Fig.~\ref{fig:eqnstateh95}
shows the equation of state for the case $h =0.95 \sigma$.  The force
$F$ appears to diverge at the packing fraction $\phi_K= \pi \sigma/(2
H_d$), which is the packing fraction of the configurations in
Fig.~\ref{fig:configs}(b) or Fig.~\ref{fig:configs}(c).  The equation
of state for the dimensionless force $f \equiv \beta F \sigma$ is
approximated fairly well by $1/f =(L-L_K)/N$, where $L_K=N \sigma/2$,
but the divergence in $f$ predicted by this approximation at
$\phi=\phi_K$ is avoided, as the curve veers off around the density
$\phi_G$: $f$ truly diverges only at the maximum possible density
$\phi_{{\rm max}}$, which corresponds to the crystalline arrangement
shown in Fig.~\ref{fig:configs}(a).  The equation of state changes its
form just above $\phi_G$, when some disks become locked into positions
away from the channel wall (as in Fig.~\ref{fig:configs}(a)).  We use
the notation $\phi_K$ as the correlation length for the zigzag order
($\xi_2$) \textit{appears} to diverge as $\phi \to \phi_K$: it grows
approximately as $\sim \exp({\rm const.}\times f)$ \cite{Godfrey:15}.
The departure of the equation of state from the straight line at the
Gardner packing fraction $\phi_G$ indicates that for $\phi > \phi_G$
the glass system has acquired new structural features.  In
Fig.~\ref{fig:density} we have plotted the average density $\rho_1(z)$
as a function of the distance from the channel wall to show the
emergence of one of these new structural features when $\phi >
\phi_G$: an increased probability of disks being found at distance
${\approx}\epsilon\sigma$ away from the wall.

In Fig.~\ref{fig:eqnstateh95} we show how the correlation length $\xi$
obtained from the logarithm of the ratio of the third eigenvalue of
the transfer matrix to the first eigenvalue varies with the packing
fraction $\phi$.  (The similar expression involving the second
eigenvalue determines the length scale $\xi_2$ of zigzag order.)  The
most striking feature of the behavior of $\xi$ as a function of $\phi$
is the peak.  The peak occurs at the packing fraction $\phi_G$ where
some of the disks start to be locked into positions at distances
$\epsilon\sigma$ away from the channel walls, as in the state of
maximum density in Fig.~\ref{fig:configs}(a) \cite{Godfrey:15}.  A
correlation length which rises to a maximum value and then falls is
typical of an avoided transition.  In Ref.~\cite{hicks:18} we
identified the peak in $\xi$ with the ``avoided'' Gardner transition
and showed that the length scale associated with the Gardner
transition defined via the overlap of two copies of the system $A$ and
$B$ was indeed described by $\xi$.  For $h=0.95 \sigma$, the peak
occurs at a packing fraction $\phi_G \approx 0.8049$, which is lower
than $\phi_K \approx 0.8055$.

\section{Dependence on channel width}
\label{epsilondep}
\begin{figure}
  \includegraphics[width=\columnwidth]{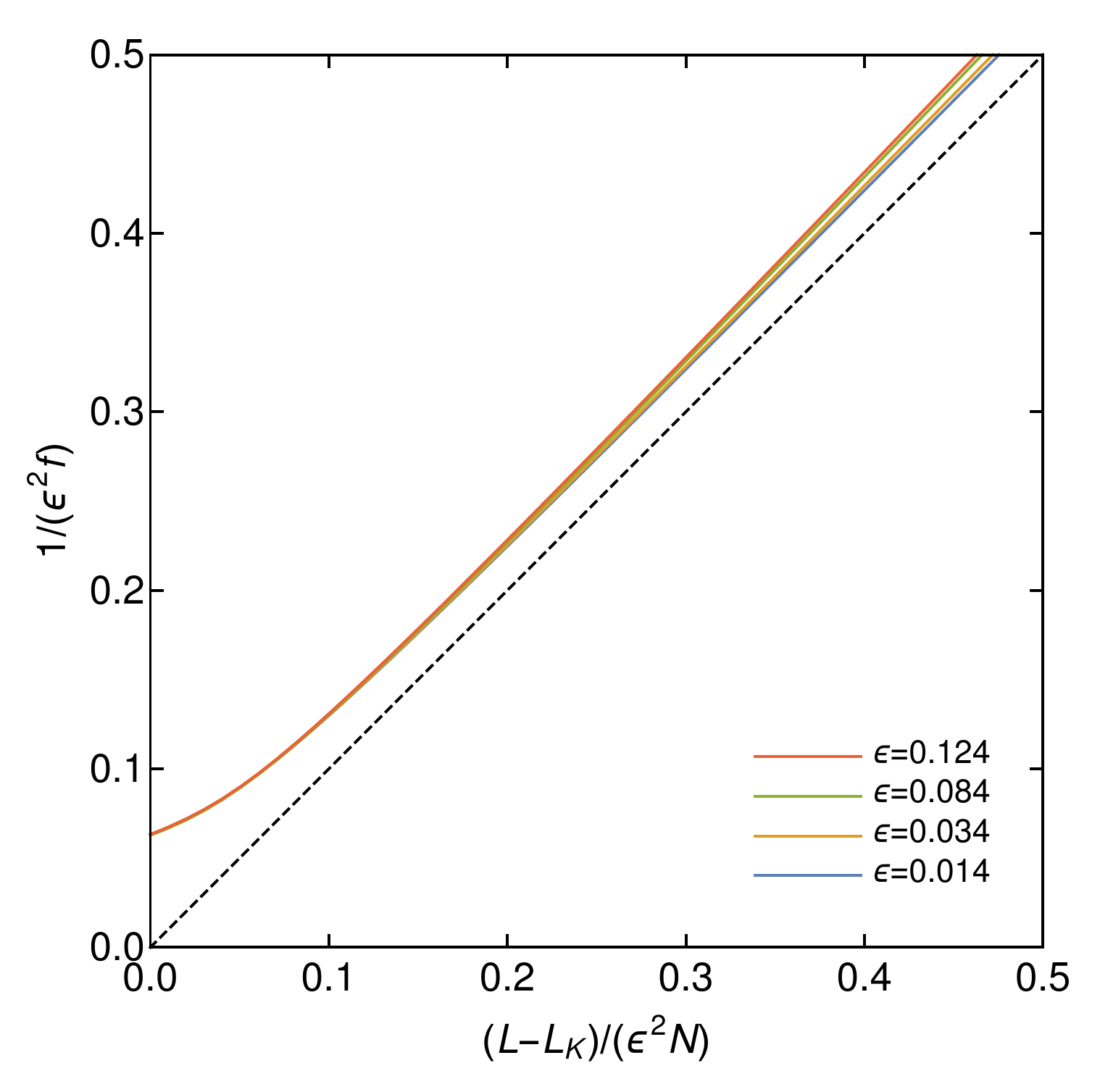}
  \caption{Plot of $1/(\epsilon^2f)$ versus $(L-L_K)/(\epsilon^2 N\sigma)$,
    where $f\equiv \beta F \sigma$ and $L_K=N \sigma/2$ is the length
    of the system in the configurations shown in
    Fig.~\ref{fig:configs}(b) or Fig.~\ref{fig:configs}(c).  The
    equation of state continues to negative values of
    $(L-L_K)/(\epsilon^2 N)$ and reaches $-1/6$ when $1/(\epsilon^2f)$
    becomes $0$, which corresponds to $\phi=\phi_{{\rm max}}$.  (We
    have so little data for the region $L < L_K$ that we have not
    plotted any of it.)  The dashed line corresponds to
    $1/f=(L-L_K)/N$.  The width of the channel is related to
    $\epsilon$ by $h=\sqrt{3}\sigma/2+\epsilon \sigma$. }
  \label{fig:scaledEOS}
\end{figure}

\begin{figure}
  \includegraphics[width=\columnwidth]{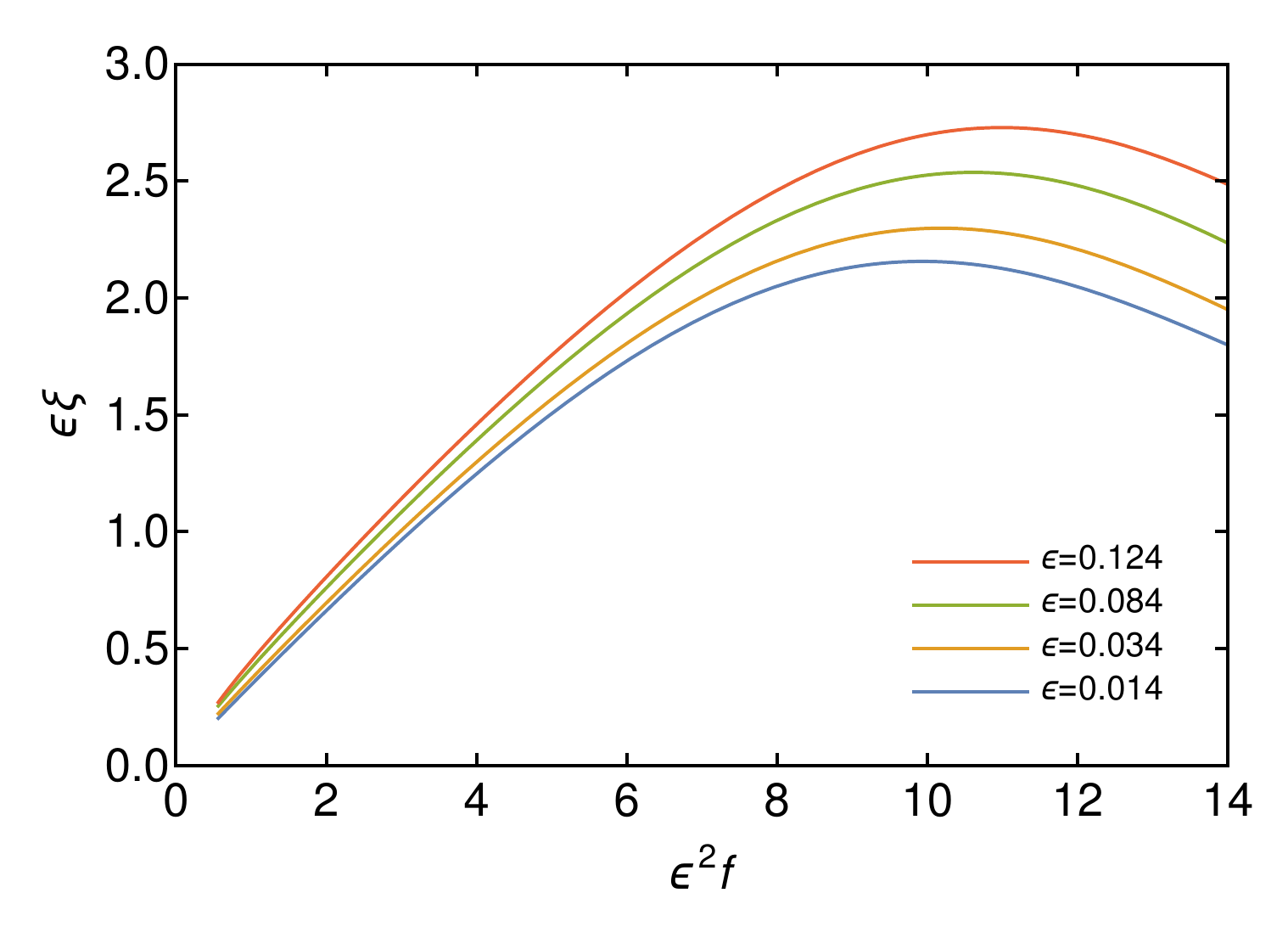}
  \caption{Plot of the scaled correlation length $\epsilon \xi$ versus
    the scaled force $\epsilon^2 f$, for various values
    of~$\epsilon$.}
  \label{fig:PTSxi}
\end{figure}
In this section we shall give results for the behavior of the equation of state and the Gardner length $\xi$ for general values of the width parameter $\epsilon$, where  $\epsilon=h/\sigma-\sqrt3/2$. Fig.~\ref{fig:scaledEOS} shows results for the equation of state in
the region $ \phi < \phi_K$ for several values of
$\epsilon$.  The dashed line is a plot of the
dimensionless force $f = \beta F\sigma$ as $1/f=(L-L_K)/N$;
Fig.~\ref{fig:scaledEOS} shows this is is accurate to order
$\epsilon^2$.  Notice that there are other shortcomings in this
approximate equation of state.  It fails when $\phi > \phi_G$ due to
the changes in the structure of the glass above $\phi_G$.

In Fig.~\ref{fig:PTSxi} we have studied $\xi$ as a function of
$\epsilon$ and the dimensionless force $f$.  If one plots $\epsilon
\xi$ against $\epsilon^2 f$, there seems to be a rather rough
``collapse'' of the data onto a universal curve.  We suspect that it
is only in the limit $\epsilon \to 0$ that the collapse becomes
convincing.  In that limit, the difference between $\phi_K$ and
$\phi_{{\rm max}}$ decreases as $\epsilon^2$, and $\phi_G$ (the
packing fraction where $\xi$ peaks as a function of $\phi$) gets
closer to $\phi_K$.  According to Fig.~\ref{fig:PTSxi}, the growth of
$\xi$ as $\epsilon \to 0$ has the approximate scaling form $\epsilon
\xi = X(\epsilon^2 f)$, with $\xi \approx \epsilon f$ when $\epsilon^2
f$ is small.  In the region where $\xi$ is large and increasing with
$f$ but $\epsilon^2 f$ is still small, we find
${\rm Var}\,x_{NN} \sim \epsilon/f$ \cite{Godfrey:15},
$B_{\infty}\sim f^2$, $G_{\infty} \sim 1/\epsilon^2$, and $\xi \to 0.5
\sqrt{B_{\infty}/G_{\infty}} \sim \epsilon f$.

On taking $\epsilon \to 0$, the avoided transition become sharper, and
more similar to a true transition.  Exactly at the peak, $f
\epsilon^2$ is of order unity, according to Fig.~\ref{fig:PTSxi}, so
that $\xi_G$, the Gardner length scale at the peak, varies as $\sim
1/\epsilon$, which shows that the Gardner length can be made
arbitrarily large by choosing $\epsilon$ to be sufficiently small.

\section{The strain-strain correlation length}
\label{strains}

\begin{figure}
  \includegraphics[width=\columnwidth]{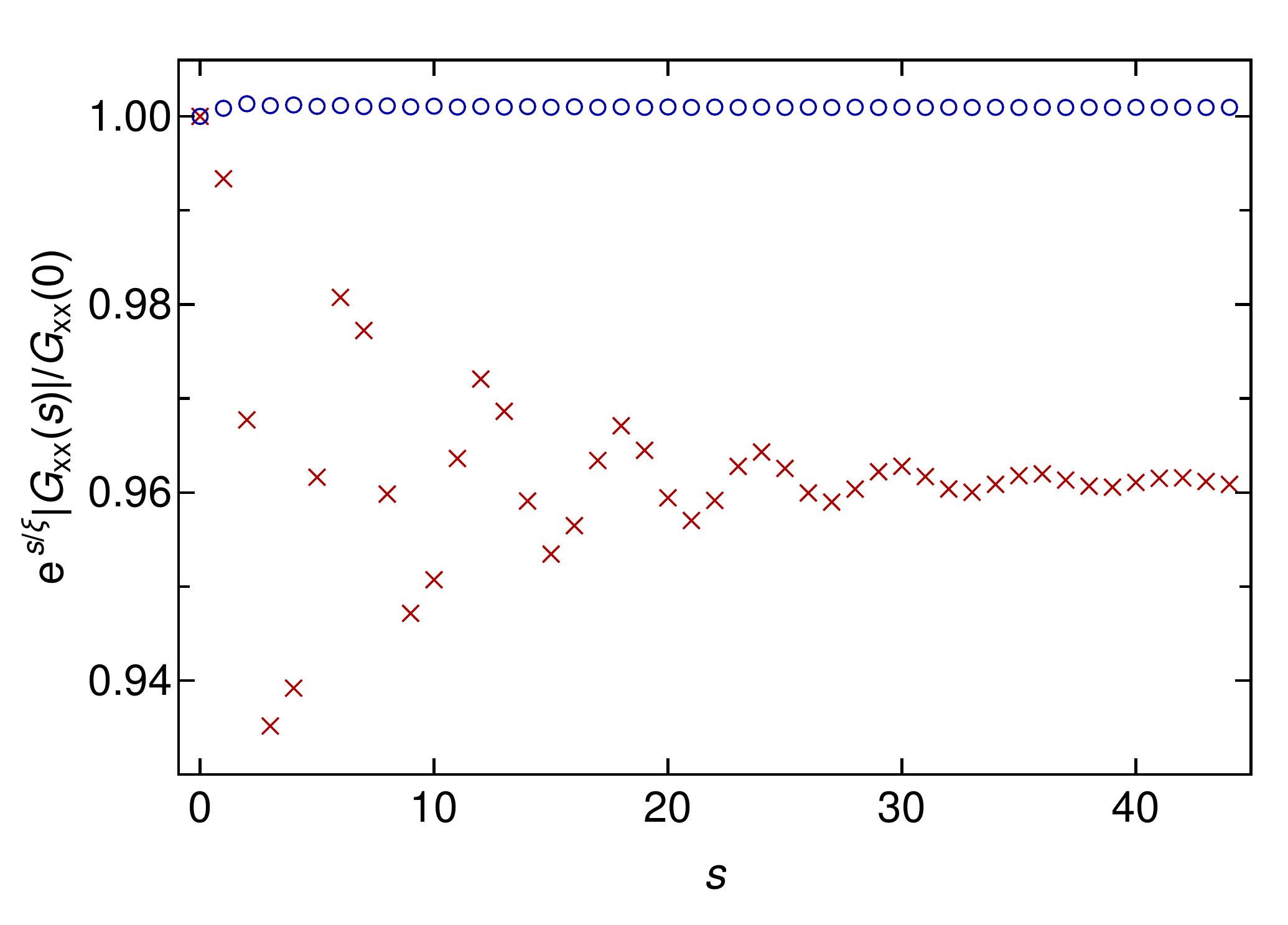}
  \caption{The ratio $\exp(s/\xi) |G_{xx}(s)| /G_{xx}(0)$ versus
    separation $s$ at two values of $f$; one below the peak in the
    $\xi$ versus $f$ plot (see Fig.~\ref{fig:PTSxi}) at $f=500$ (blue
    circles), where $\xi\approx 14.9$ and one above beyond the peak at
    $f=2500$ (red crosses), where $\xi \approx 19.5$ for the width
    $h=0.95 \sigma$.  If Eq.~(\ref{strain}) were a perfect fit, the
    ratio would be unity for all $s$. }
  \label{fig:expdecay}
\end{figure}

We have  studied the strain-strain correlation function
$G_{xx}(s)$ defined by $G_{xx}(s) \equiv \langle \tilde{x}_i
\tilde{x}_{i+s}\rangle$.  Here $\tilde{x}_i \equiv
(x_{i+1}-x_i-\langle x_{i+1}-x_i\rangle)$ is the $x$-component of the
strain between particle $i+1$ and particle $i$.  Note that $\langle
\tilde{x}_i \rangle=0$.  The average spacing between the disks along
the $x$ axis is $\langle x_{i+1}-x_i\rangle=L/N$.
Fig.~\ref{fig:expdecay} shows that $G_{xx}(s)$ decays with $s$ on the
Gardner length scale $\xi$ for large $s$ at any packing fraction
$\phi$ and that for $\phi < \phi_G$ it is accurately described by the
equation
\begin{equation}
  G_{xx}(s)  \approx (-1)^s {\rm Var}\,x_{NN} \exp(-s/\xi).
  \label{strain}
\end{equation}
${\rm Var}\,x_{NN}$ denotes the variance of the nearest-neighbor
spacing $x_{i+1}-x_i$ and equals $G_{xx}(0)$ \cite{Godfrey:15}.  The
oscillation in sign arises from the fact that the two rows of disks
tend to move independently of each other so that if one
nearest-neigbor distance is increased, the following one will
decrease.  $\xi$ in Eq.~(\ref{strain}) is equal to that calculated
from the logarithm of the ratio of the third eigenvalue of the
transfer matrix to the first eigenvalue of the transfer matrix.

Notice that studying the strain-strain correlation function can be
done within a single copy of the system.  When using the transfer
matrix technique, it is simple to determine the strain-strain
correlation function.  However, when doing a molecular dynamics
simulation in higher dimensions it might be easier to study the
stress-stress correlation function.  It seems natural to expect that
both types of correlation function will decay exponentially on the
same length scale $\xi$.

For understanding the behavior of the strain-strain correlation
function at densities $ \phi < \phi_G$, we shall find it useful to
introduce an effective Hamiltonian.  Our system has the feature that
the disks cannot pass each other, so their ordering persists forever:
they are ``caged'' even in the low density region.  (In three
dimensions caging is only effective on time scales less than the alpha
relaxation time.)  With the effective Hamiltonian
\begin{equation}
  \mathcal{H} = \frac{1}{2} B_{\infty} \sum _n(\tilde{x}_n+\tilde{x}_{n+1})^2 +\frac{1}{2} G_{\infty}
  \sum_n (\tilde{x}_n-\tilde{x}_{n+1})^2,
  \label{effH}
\end{equation}
Eq.~(\ref{strain}) is recovered as an equality, provided we choose
$B_{\infty}$ and $G_{\infty}$ so that
\begin{equation}
  \xi=\frac{1}{2\, {\rm arctanh}(\sqrt{G_{\infty}/B_{\infty}})},
  \label{xiBg}
\end{equation}
and ${\rm Var}\,x_{NN}=1/(4 \beta
\sqrt{B_{\infty}G_{\infty}})$.  The first term in brackets in
Eq.~(\ref{effH}) involves $(\tilde{x}_n+\tilde{x}_{n+1}) \equiv
x_{n+2}-x_n-2 L/N$, while the second term involves
$(\tilde{x}_n-\tilde{x}_{n+1}) \equiv 2 x_{n+1}-x_n-x_{n+2}$.  Because
of the zigzag order, the first term is the spacing between disks in
the same row.  This separation is actually on average smaller than
that between disks $n$ and $n+1$ \cite{Godfrey:15}.  If the pistons
are pushed in, this elastic term measures the free energy cost of
compressing the system and the coefficient $B_{\infty}$ is the bulk
modulus of the system.  The second term represents the free-energy
cost of a shear displacement in which the disks in the top row slide
with respect to those in the bottom row, as shown in
Fig.~\ref{fig:configs}.  This can be recognized from the fact that it
is non-zero when $x_{n+1}$ which is in (say) the top row, moves away
from the mid-point of the two disks $n$ and $n+2$ in the bottom row,
$(x_{n}+x_{n+2})/2$, as in Fig.~\ref{fig:configs}(b).  Its coefficient
will be the shear modulus $G_{\infty}$ of our system.  The term
represents an effective coupling, as the entropy of the system is
greatest when the disks in one row hover above the dips between
adjacent disks in the other row; thus, it is a many-body rather than
just a pairwise effective interaction.

Equation~(\ref{effH}) will have corrections to it that involve higher
powers of $\tilde{x}_n$.  At densities above $\phi_G$,
Fig.~\ref{fig:expdecay} shows that $G_{xx}(s)$ has features for $s <
\xi$ which are not captured in the simple exponential form of
Eq.~(\ref{strain}) predicted by Eq.~(\ref{effH}), but this form
remains accurate to ${\approx} 96\%$ at the highest density studied,
which corresponds to $f=2500$.  The departures from the simple
exponential decay are consequences of the structural changes that
emerge for $\phi>\phi_G$.

Thus the effective Hamiltonian of Eq.~(\ref{effH}) provides a good
description of the system for $\phi < \phi_G$.  We shall interpret it
in two ways.  The first is to introduce a new variable, the ``tilt''
$t_i=\tilde{x}_i/{\rm sign}(y_i)$.  The ${\rm sign}(y_i)$ term removes the
$(-1)^s$ sign oscillation in Eq.~(\ref{strain}), so that $\langle t_i
t_{i+s}\rangle \sim \exp(-s/\xi)$, a simple exponential decay.  In
terms of the variables $t_n$, the effective Hamiltonian is
\begin{equation}
  \mathcal{H} = \frac{1}{2} B_{\infty} \sum _n(t_{n+1}-t_{n})^2 +\frac{1}{2} G_{\infty}
  \sum_n (t_{n+1}+t_n)^2.
  \label{effHtn}
\end{equation}
Note that if one is working in the region where $B_{\infty} \gg
G_{\infty}$, the first term is minimized by having the variables $t_n$
of the same sign.  Furthermore, if the variation of $t_n$ with $n$ is
small one can approximate $t_{n+1}-t_n \to \partial t(x)/\partial x$
and $t_{n+1}+t_n \to 2 t(x)$ so that the continuum version of
Eq.~(\ref{effHtn}) is
\begin{equation}
  \mathcal{H} =\int \,dx \left[ \frac{1}{2} B_{\infty} (\partial t(x)/\partial x)^2 +\frac{4}{2} G_{\infty} t(x)^2\right].
  \label{effHtx}
\end{equation}
From Eq.~(\ref{effHtx}) it is easy to see that there will be a
correlation length $\xi \to 0.5 \sqrt{B_{\infty}/G_{\infty}}$.

The second way of understanding the origin of the length scale $\xi$
will be useful in higher dimensions.  It is to Fourier transform the
strain-strain correlation function of Eq.~(\ref{strain}), assuming
that the particle are on average spaced by $a =L/N$.  The result for
the Fourier transform $\tilde{G}_{xx}(k)$ for $\xi$ large is
\begin{equation}
  \tilde{G}_{xx}(k)= \frac{T}{4G_{\infty}+2 B_{\infty} [1+\cos(k a)]}.
  \label{FTstrain}
\end{equation}
For $k a =\pi+q a$ with $qa \ll 1$, the denominator is proportional to
$(1/\xi^2+(qa)^2)$.  Thus the long length scale $\xi$ is associated
with a softening of the elastic modes at the ``zone boundary'', $k =
\pi/a$.  It is the many-body term involving $G_{\infty}$ in
Eq.~(\ref{effH}) which stabilizes the system against the free sliding
of one row with respect to the other row.

\section{Discussion}
\label{discussion}

We will now discuss how this argument might extend to glasses in dimensions $d= 2$ and $d=3$.  On
time scales less than the alpha relaxation time the particles in the glass are
caged.  Effective Hamiltonians similar to that of Eq.~(\ref{effH}) can
be written down for the glass state in higher dimensions, as that
effective Hamiltonian is just one of the many possible
finite-difference forms that reduce to continuum elasticity theory in
the long-wavelength limit.  Furthermore, in the region near the
J-point, the number of nearest-neighbors will be close to $2d$ on
average, rather as on a simple-cubic lattice for $d =3$ or a square
lattice in $d=2$ \cite{epitome,brito:06,LiuNagel,Ozawa:17c}.  A
simple-cubic or a square lattice with only pairwise interactions
between nearest-neigbors on the lattice is unstable against sliding
one plane (or row) of particles with respect to the rest.  This kind
of instability is at the wavevector $k=\pi/a$, where $a$ is the
lattice spacing.  It is the analogue of sliding the top row of disks
with respect to the bottom row in the disks in the channel system.
Indeed, as in our channel system, it is only the existence of
next-nearest-neighbor interactions which makes such cubic systems
stable and gives rise to a non-zero shear modulus $G$.  In glasses the
situation will be more complicated, as there will be no freely sliding
planes of spheres; in fact, no well-defined planes of spheres.
However, near the J-point a snapshot of the system would show that it
is close to a state of marginal stability; $G$ is actually zero at the
J-point itself.  It seems natural to expect that at wave-vectors
$|\mathbf{k}|$ that correspond to a displacement similar to that near
$\pi/a$, there will be a peak in the strain-strain correlation
function, as in Eq.~(\ref{FTstrain}) due to the expected softening at
such a wavevector.  In fact, such a peak may have already been
observed in the stress-stress correlation functions studied in
Ref.~\cite{wu:17} for the case of a truncated Lennard-Jones potential.
In that work, no peak was visible in states of a lower packing
fraction $\phi=0.699$, but there was a striking peak for states with a
packing fraction $\phi=0.80$, which is closer to the J-point of this
system.  The length scale $\xi$ can be extracted from the peak in the
strain-strain correlation function (the higher dimensional analogue of
Eq.~(\ref{FTstrain})) and provided $B_{\infty}\gg G_{\infty}$ it
should vary as $\sim \sqrt{B_{\infty}/G_{\infty}}$. Alas, simulations
on hard or soft spheres have not yet been sufficiently developed to
allow this relationship to be investigated for three dimensional
systems \cite{ berthier:19}.

It has not escaped our notice that the softening as the J-point is
approached will contribute to the ``boson peak" effect
\cite{silbert:09}: a boson peak arises when there are more low
frequency phonon modes than might have been expected from just the
\textit{long-wavelength} elastic modes included in the Debye
approximation.

For soft spheres the J-point \cite{scalliet:19b,scalliet:19c} has been shown to lie within the Gardner phase.  We believe that the Gardner transition is an avoided transition in physical dimensions so we suspect that that Gardner-like features will only be seen at temperatures and packing fractions which are close  to the J-point, (which is a $T= 0$ point). Support for this idea and also  our formula $\xi \sim \sqrt{B_{\infty}/G_{\infty}}$ comes from the work of Refs.~\cite{DeGiuli:14b,DeGiuli:14}. There it was found that for packing fractions larger than $\phi_J$, there was a  length scale in jammed states of soft spheres which has  had the same dependence on the elastic moduli as in this paper. This length scale marked the crossover in the response to a force dipole between the fluctuation-dominated near field and the elastic-continuum far-field. This length is also the scattering length associated with the "anomalous modes" that are important near the J-point.  Dynamical criticality  at finite temperatures near the jamming transition has been extensively discussed in \cite{ikeda:13, degiuli:15}.

\section*{Acknowledgments} 
  We thank Ludovic Berthier, and Camille Scalliet for useful exchanges,
   Kamran Karini, Craig Maloney and Steve Teitel for information on
  the work in Ref.~\cite{wu:17}, and Eric De Giuli for pointing out the features mentioned in the previous paragraph.

\section*{References}
\bibliography{refs}
\end{document}